\documentclass[epj]{svjour}
\usepackage{graphics}
\usepackage{hyperref}
\hypersetup{colorlinks = true, allcolors = blue}
\usepackage[utf8]{inputenc}
\usepackage{authblk}
\usepackage{hepparticles}
\usepackage{hepunits}
\usepackage{hepnames}
\usepackage[dvipsnames]{xcolor}
\usepackage[demo]{graphicx}
\usepackage{caption}
\usepackage{subcaption}

\newcommand{\hssb}{$\textrm{H}\rightarrow \textrm{s} \bar{\textrm{s}}$}
\newcommand{\hcc}{$\textrm{H}\rightarrow \textrm{c} \bar{\textrm{c}}$}
\newcommand{\hbb}{$\textrm{H}\rightarrow \textrm{b} \bar{\textrm{b}}$}
\newcommand{\hgg}{$\textrm{H}\rightarrow \textrm{g} \textrm{g}$}

\begin{document}

\title{Higgs and top physics reconstruction challenges and opportunities at FCC-ee}

\author{Patrizia Azzi\inst{1,2} \and Loukas Gouskos\inst{2} \and Michele Selvaggi\inst{2} \and Frank Simon\inst{3} 
}                     

\institute{INFN, Sezione di Padova, Padova, Italy \and  CERN, EP Department, Geneva, Switzerland \and Max-Planck-Institut f\"ur Physik, M\"unchen, Germany
}

\authorrunning{Azzi, Gouskos, Selvaggi, Simon}
\titlerunning{Higgs and top physics challenges}
\date{Accepted November 26, 2021}

\abstract{
The Higgs bosons and the top quark decay into rich and diverse final states, containing both light and heavy quarks, gluons, photons as well as W and Z bosons. This article reviews the challenges involved in reconstructing Higgs and top events at the FCC-ee and identifies the areas where novel developments are needed. The precise identification and reconstruction of these final states at the FCC-ee rely on the capability of the detector to provide excellent flavour tagging, jet energy and angular resolution, and global kinematic event reconstruction. Excellent flavour tagging performance requires low-material vertex and tracking detectors, and advanced machine learning techniques as successfully employed in LHC experiments.  In addition, the Z pole run will provide abundant samples of heavy flavour partons that can be used for calibration of the tagging algorithms. For the reconstruction of jets, leptons, and missing energy, particle-flow algorithms are crucial to explore the full potential of the highly granular tracking and calorimeter systems, and give access to excellent energy--momentum resolution and precise identification of heavy bosons in their hadronic decays. This enables, among many other key elements, the reconstruction of Higgsstrahlung processes with leptonically  and hadronically decaying Z bosons, and an almost background-free identification of top quark pair events. Exploiting the full available kinematic constraints together with exclusive jet clustering algorithms will allow for the optimisation of global event reconstruction with kinematic fitting techniques.
%
} 
\maketitle

\section{Introduction}
\label{section:intro}
Precision measurements of the Higgs boson and of the top quark are key objectives of the FCC-ee physics programme \cite{Abada:2019lih}, and define the requirements for the detector design and the reconstruction algorithms for the higher-energy stages of the machine at $\sqrt{s}=$\,240 GeV and 350/365 GeV. The four and six-fermion final states of HZ and $t\bar{t}$ events are typically characterised by the presence of multiple hadronic jets, heavy quarks, leptons, and neutrinos. To fully capitalise on the potential for precision measurements provided by the high integrated luminosity and the clean experimental environment at the FCC-ee, specific detector technologies as well as reconstruction strategies and algorithms are needed to achieve an optimal performance for the complex Higgs boson and top quark events. Jet energy resolution, flavour tagging in hadronic jets, and global event and missing energy reconstruction are the key ingredients to achieve this goal. By taking advantage of the knowledge of the initial state, powerful kinematic constraints can be applied, which have the potential to significantly increase the precision of the overall event reconstruction as well as that of individual objects. 

In the context of Higgs boson physics, the quality of jet energy and directional reconstruction enters at various levels, from the quality of the identification of Higgsstrahlung events for hadronic Z boson decays, to the precision of the measurement of the Higgs boson mass and to the identification of distinct decay modes. Flavour tagging is central for the latter, given the dominance of hadronic decays of the Higgs boson. A highly efficient discrimination of $b$, $c$, and gluon jets allows to access novel decay modes that cannot be identified at the LHC, adding qualitatively new dimensions to the Higgs physics programme.

The identification of the complex events of top quark pair production profits significantly from an accurate jet and global event reconstruction via an excellent measurement of the invariant mass of gauge bosons occurring in the decay of top quarks and in background processes, and from an highly precise measurement of the invariant mass of the overall final state. Also here, flavour tagging is crucial, since top quark pair events are characterised by the presence of two $b$ quark jets. Their clean identification enables improved signal and background separation. 

In the following, we discuss key reconstruction techniques that are being developed to achieve the goals for Higgs boson and top quark physics at lepton colliders. These are particle flow algorithms for physics object and overall event reconstruction, and advanced flavour tagging algorithms. Following this, the potential and future opportunities are discussed with selected examples. Since studies for FCC-ee are still in early stages in many cases, concrete examples are also taken from linear collider-based studies, which have a high degree of overlap in the physics goals, very similar experimental conditions, and provide starting points for setting goals for FCC-ee detector concepts, reconstruction and analysis strategies. In this context, full-simulation studies of detector performance and selected physics channels performed in the context of CLIC at centre-of-mass energies of 350 GeV and 380 GeV and in the context of ILC at 250 GeV are particularly relevant for the higher-energy stages of FCC-ee at 240 GeV, at the top quark pair threshold around 350 GeV, and at 365 GeV. 

\section{Particle flow and beyond : reconstructing individual particles, jets, and global event kinematics}
\label{sec:PFA}

Particle flow (PF) aims to achieve the best possible reconstruction of individual final-state particles by optimally combining the information available from the different subsystems of the detector to achieve a global particle-level description of reconstructed events. Besides the clear benefits for jet energy reconstruction provided by improved single-particle measurements, this enables more performant heavy flavour tagging of jets, and the exploitation of kinematic constraints by global event kinematic fitting with advanced techniques, expanded further in the two following sections. 
A first generation of PF was developed for the ALEPH experiment at LEP~\cite{Buskulic:1994wz}, which enabled that experiment to achieve the best jet energy resolution of all four LEP experiments. However, the performance of this algorithm was limited by the relatively low granularity of the calorimeters, allowing the detection of neutral hadrons only via a significant excess of calorimeter energy compared to the sum of particle track momenta in the same geometric region. The CMS experiment at LHC has developed a sophisticated algorithm that results in significant improvements of the jet energy resolution compared to calorimeter-only reconstruction \cite{Sirunyan:2017ulk}. 

The detector concepts for the linear $e^+e^-$ colliders ILC and CLIC are specifically designed for the optimal performance of PF algorithms, achieving a jet energy resolution of 3\%--4\% over a wide energy range with sophisticated reconstruction algorithms \cite{Thomson:2009rp,Marshall:2012ry,ILD:2020qve}, as required for the efficient separation of $W$ and $Z$ bosons in hadronic final states. The crucial detector feature to achieve this performance is a high spatial granularity in the calorimeter systems \cite{Sefkow:2015hna}, which is exploited in the pattern recognition required to achieve an accurate matching of subdetector information, and also provides the potential for improved calorimetric reconstruction using software compensation methods \cite{Adloff:2012gv,Tran:2017tgr}. This principle is also adopted for the CLD detector concept \cite{Bacchetta:2019fmz} for FCC-ee, which is derived from the CLIC detector design, adjusted for the different experimental conditions in terms of collision energy and operation mode. 

The accurate reconstruction of hadronic jets is of importance in many areas of Higgs boson and top quark physics due to the prevalence of hadronic final states in their decay. In general, improved performance in this area results in reduced backgrounds and higher signal significance or more accurate measurements. Comprehensive studies have been performed in the context of CLIC, which demonstrate the capabilities of a particle-flow-optimised detector, albeit for an energy range largely beyond that achievable at FCC-ee~\cite{Abramowicz:2016zbo,Abramowicz:2018rjq}. Similarly, studies in the context of ILC show the potential provided by the accurate reconstruction of top quark pair production events, and by the precise reconstruction of jet directions for the measurement of the Higgs boson mass in HZ events with H $\rightarrow$ $b\bar{b}$ and Z decaying into quark pairs \cite{ILD:2020qve}. 

While the main motivation for PF, and for detector concepts optimised for its application, is generally given by the jet energy resolution, the benefits of this concept extend far beyond that one metric alone. PF algorithms improve background rejection, as implemented for CLIC in particular for the higher-energy stages. The additional precision of the reconstruction at the level of individual particles also supports more powerful flavour tagging of reconstructed jets, as discussed further in section \autoref{sec:Flavor}, and provides the potential for flavour tagging strategies exploiting the full event information. The information provided by a global particle level PF-based event reconstruction also allows for a precise reconstruction of global event kinematics, which can be used together with the constraints provided by the well-known initial state of the collision in kinematic fits to significantly improve the measurement of kinematic variables. In the context of Higgs boson and top quark physics, this is for example beneficial for the rejection of physics background in hadronic and semi-leptonic final states, for the kinematic reconstruction of particle masses and for the constraints on the jet energy scales, as shown for top quark pair events in \cite{Seidel:2013sqa}. It is expected that significant additional improvement on several aspects of jet and event reconstruction with PF can be achieved with machine learning techniques that use the available information on various levels: from low-level object reconstruction, to jet finding, to global event reconstruction aspects such as jet pairing and other key aspects of precision measurements of the complex final states characterising many Higgs boson and top quark events at the FCC-ee.

\section{Flavour tagging: separating bottom and charm quarks, and gluons}
\label{sec:Flavor}

The identification of the flavour of the parton that initiated the formation of a jet, known as flavour tagging, is a crucial ingredient to determine the complex final states that result from Higgs boson and top quark decays. The measurement of the     coupling of the Higgs boson to quarks and gluons directly relies on the ability to identify the flavour of these jets. One of the highest priorities of the FCC-ee physics programme is to measure with O($\sim 1\%$) or better precision the coupling of the Higgs boson to bottom and charm quarks, and gluons. Moreover, measurements of top quark properties also crucially rely on the identification of bottom quarks. Yet, to achieve the desired precision requires advancements in the detector design and on the flavour tagging algorithms. 

Jet flavour identification algorithms were already developed at LEP~\cite{Abdallah:2002xm,Proriol:1950599}. The cornerstone of these algorithms are tracks with significant displacement from the beam axis, which originate from the large lifetime of the weak decays, $\sim$1.5 ($\sim$0.4--1.0) ps for B (C) hadrons, as opposed to ``light'' hadrons that are produced with a vanishing displacement. Such a property of the tracks can be used to identify possible displaced ``secondary'' vertices (SV) in the event. Therefore, the precision at which the track impact parameter can be measured directly translates into the discrimination power between heavy and light flavour jets. For future detector designs, vertex and tracking detectors with high granularity and as little material as possible, are necessary. In addition, constraints related to the beam pipe are parameters that have to be carefully studied.  Ideally, the first layer of the vertex detector would be placed within the beam  pipe. However, this approach requires  shielding to reduce noise and interference from the beam current, which results in an increased  material budget for the signal readout of this detector. An alternative approach that can be explored is to investigate different beam pipe sizes allowing to reduce the distance between the inner tracking layer and the interaction point. 

Additional information such as the track multiplicity, the mass of the SV, and potentially the presence of a non-isolated lepton from semi-leptonic heavy flavour decays, were also some of the main ingredients of the flavour tagging algorithms. These algorithms were based on either simple selections on a combination of these SV properties, or, in more advanced implementations, exploited likelihood ratio methods using these properties. The first generation of jet flavour tagging algorithms at the LHC, e.g.\ in Refs.~\cite{CMS-PAS-BTV-11-004,ATLAS-CONF-2010-041,ATLAS-CONF-2010-042,ATLAS-CONF-2010-070,ATLAS-CONF-2010-091} were based on very similar principles, yet applied under the harsher data taking conditions of the LHC. However, taking advantage of the developments in computing (e.g.\ graphics processing units) and the availability of very large Monte Carlo simulated and collision data samples, a new era in jet flavour identification has begun. These algorithms explore directly much lower-level information (e.g.\ reconstructed particles from the PF algorithm or even reconstructed hits), which are processed using state-of-the-art machine learning (ML) algorithms, tailored to the jet tagging problem. The full spatial and energy distribution can be exploited in jet flavour identification, particularly in the separation between quark and gluon jets~\cite{CMS-DP-2017-027}. 

Finally, the nature of each of the jet constituents, or particle identification (PID), is expected to provide another useful handle in discriminating between different jet species. Powerful particle identification capabilities based on ionisation energy loss (via dE/dx or cluster counting), or via precise time-of-flight measurements, could be highly beneficial for jet flavour tagging at the FCC-ee allowing the full jet content information to be exploited. 

\begin{figure}[!h]
\centering
\includegraphics[width=0.32\textwidth]{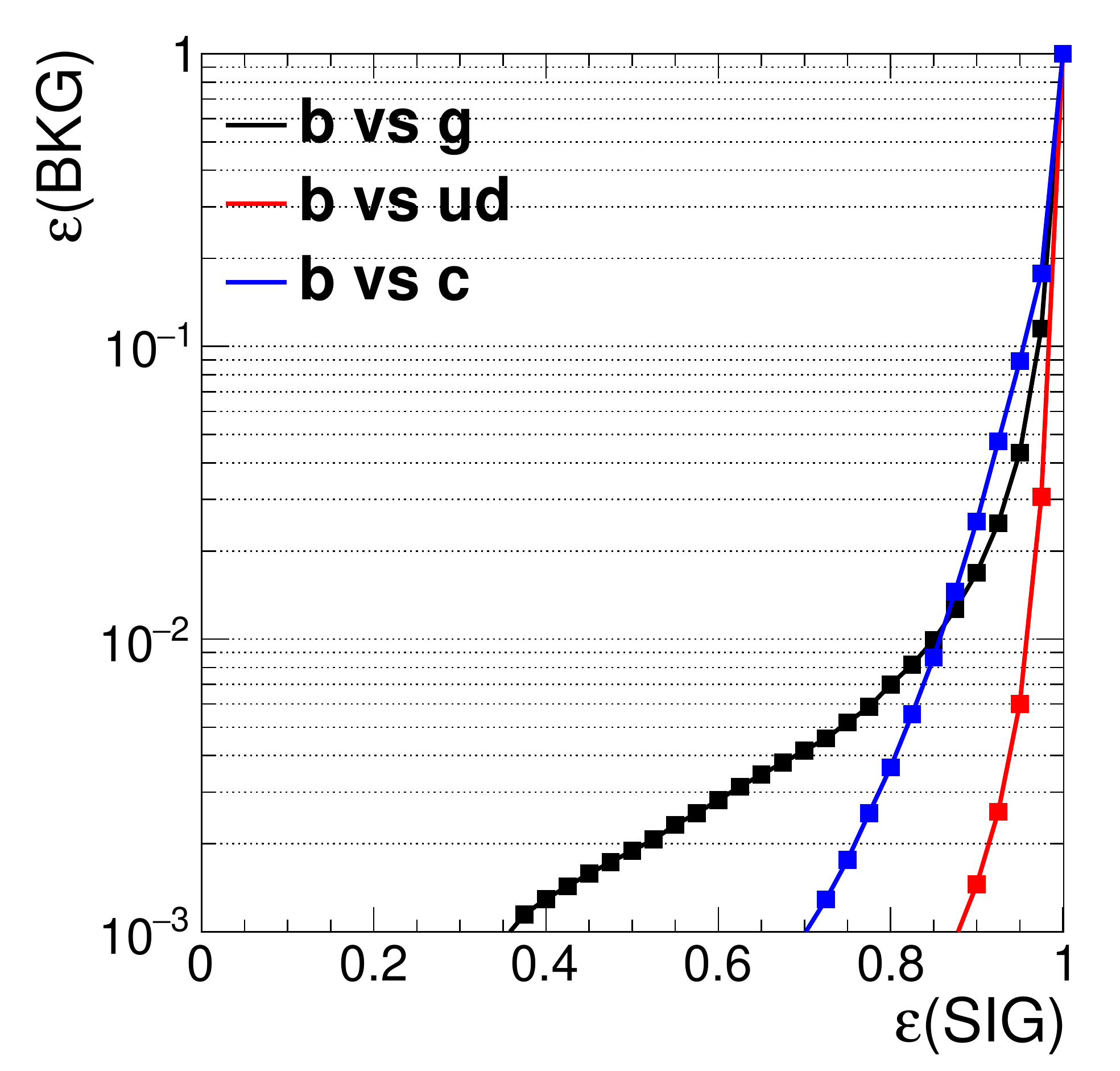}
\includegraphics[width=0.32\textwidth]{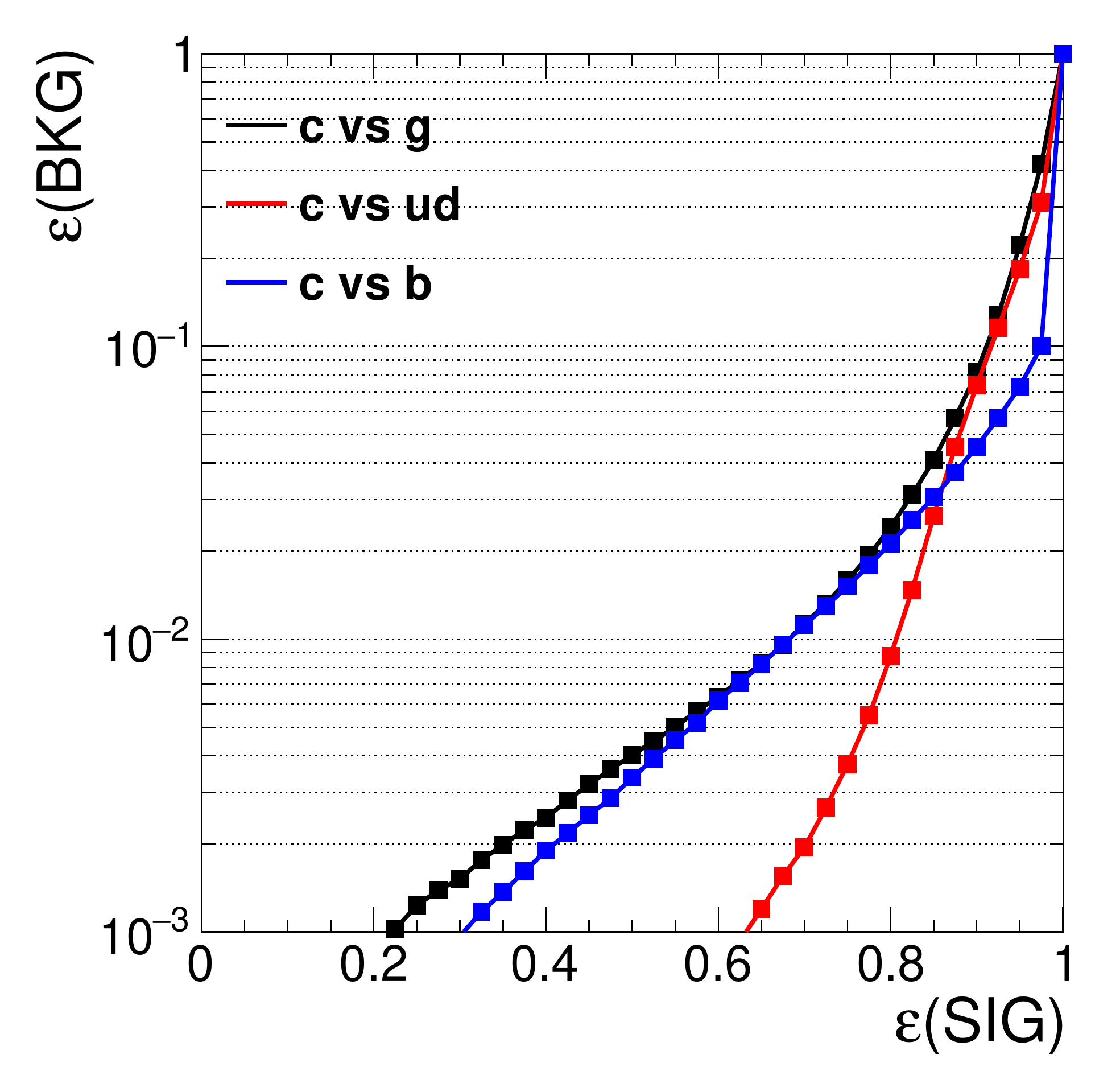}
\includegraphics[width=0.32\textwidth]{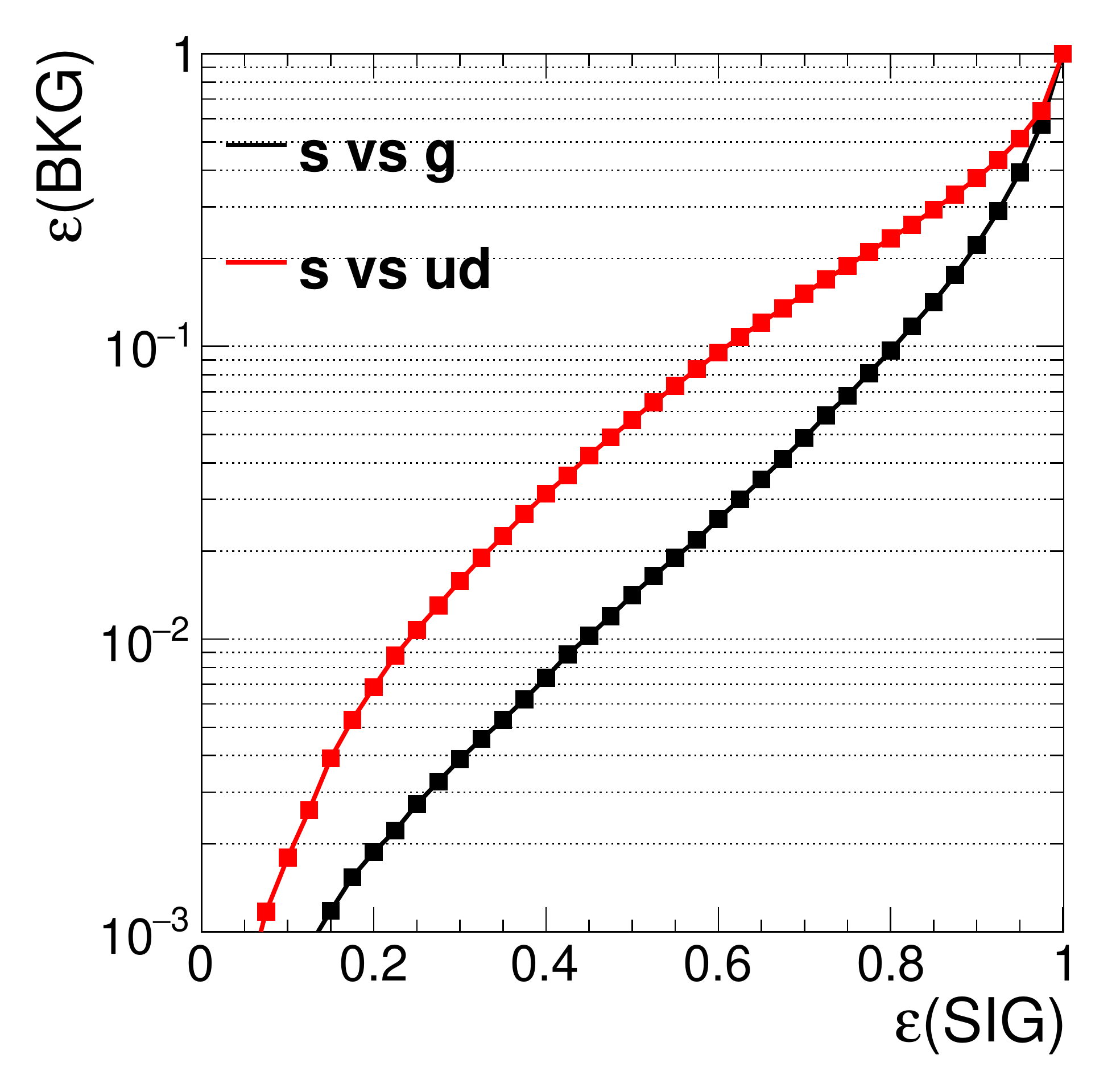}

\caption{Background contamination efficiency as a function of jet-tagging efficiency for bottom (left), charm (centre) and strange (right) quark jets in HZ events at a centre-of-mass energy of 240 GeV in the IDEA detector.}
\label{fig:fccee_dnn_flavour tagging}
\end{figure}

At the LHC, these new approaches can bring more than an order of magnitude improvement in background rejection compared to the traditional approaches~\cite{CMS:2020poo}. They are now also being explored in the context of the FCC-ee, leading to the development new flavour tagging algorithms which aim to explore the full potential of the detectors and the significantly more benign FCC-ee environment. A recent development~\cite{fftforEE} explores directly the particles reconstructed by the PF algorithm, which are then used as input to ML based flavour tagging algorithms. The algorithm is based on ParticleNet~\cite{Qu:2019gqs}, a novel approach using state-of-the-art jet representation and a neural network (NN) architecture. Preliminary results of the performance of the algorithm are displayed in \autoref{fig:fccee_dnn_flavour tagging} in terms of receiver operating characteristic (ROC) curves, for bottom and charm quark tagging. Different jet flavours from hadronic Higgs decays are taken from a Higgsstrahlung sample at $\sqrt{s} = 240$~GeV. The IDEA~\cite{Abada:2019zxq}  detector concept was used and the detector's response is simulated using the Delphes~\cite{deFavereau:2013fsa} fast detector simulation framework. Compared to previous flavour tagging algorithms, e.g., in Ref.~\cite{ILD:2020qve}, a significant improvement in performance is observed. 

Another important aspect that is expected to further improve the physics performance is the development of flavour tagging algorithms with the ability to simultaneously discriminate between multiple different jet flavours. For example, in the \hcc\ process where the  \hbb\ and \hgg\ background processes are copiously produced, this is expected to bring important improvement. Having this in mind, the jet flavour algorithm under development provides a probability for each jet to originate from different particle species, namely, up or down, strange, charm, or bottom quark, or from a gluon. In addition, this approach gives the opportunity to explore the potential of strange quark tagging in virtually unexplored physics channels, such as \hssb\ or flavour changing neutral currents (FCNC) in the top sector. Further improvements can be obtained when PID information is also included in the algorithm. PID is already exploited at the LHC to provide powerful discrimination between pions, kaons, and protons~\cite{Hines:2011uf,Powell:1322666,Giammanco:1095044}. Particularly for strange quark tagging, exploring such handles yields in an improvement in the discrimination power against first-generation quark jets. \autoref{fig:fccee_dnn_flavour tagging} (right) shows preliminary results on strange jet tagging that indicate that a powerful discrimination against up, down and gluon jets can be obtained with the use of PID. Further studies will explore the impact of timing information on PID and subsequently on jet-tagging performance.  
Finally, apart from the clear challenges in the detector and algorithm design, another big challenge is the calibration of such algorithms with the required precision in order to achieve relative uncertainty for most of the Higgs coupling measurements better than 1\%. The FCC-ee running scenario includes the operation at the Z pole with the goal to collect O($10^{12}$) events. This will provide excellent conditions for the calibration of the jet flavour tagging algorithm with unprecedented precision.

\section{Exploitation for precision measurements and opportunities for further development}

The physics potential of the techniques outlined above has already been studied for Higgs boson and top quark physics, primarily in the linear collider context. In the following, we discuss one concrete example, and then highlight areas where we see significant potential for further developments that address key remaining challenges. 

\begin{figure}
    \centering
    \includegraphics[width=0.6\textwidth]{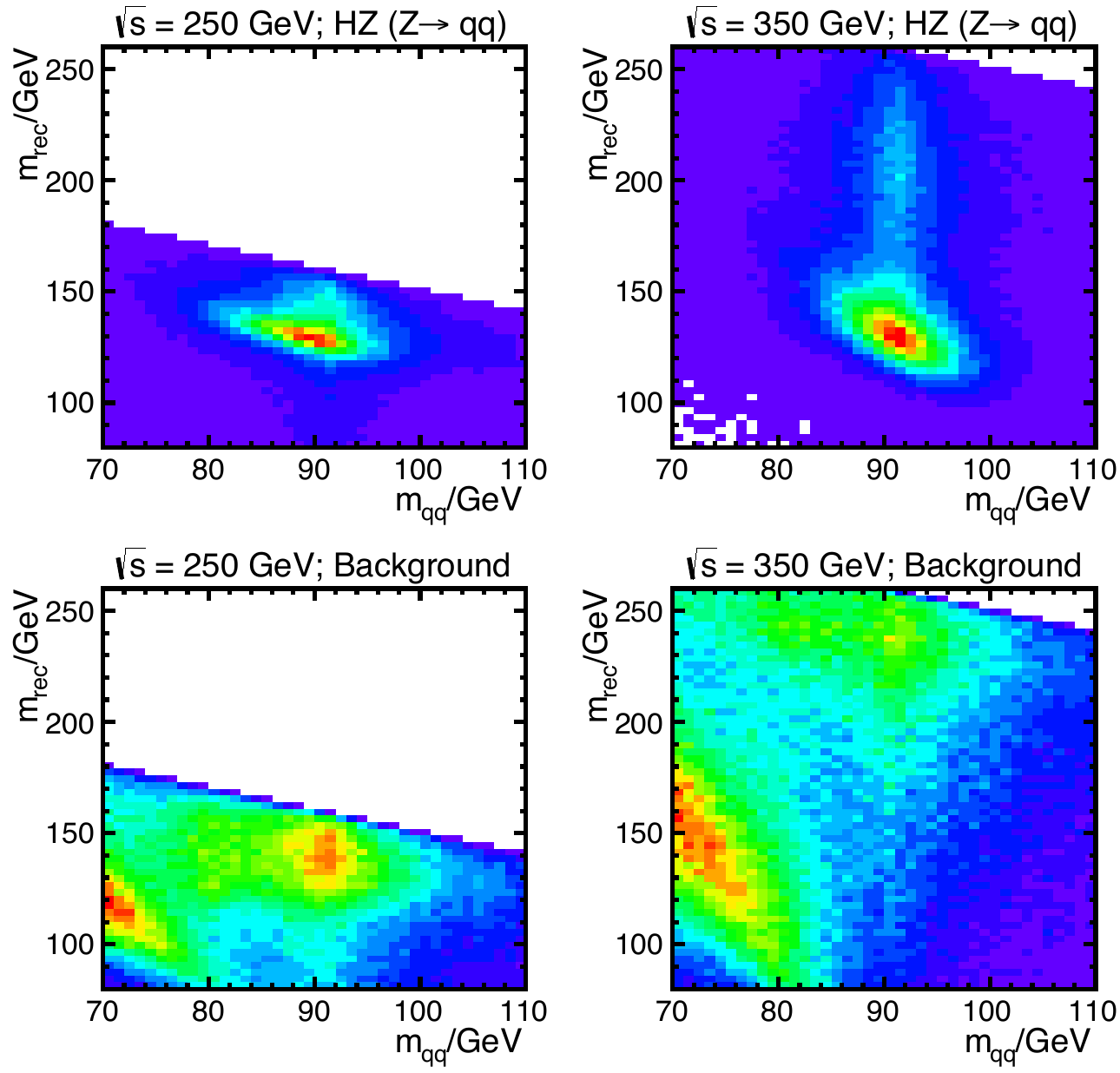}
    \caption{Reconstructed di-jet Z candidate mass, $m_{qq}$ versus reconstructed hadronic recoil mass $m_{rec}$ for a centre-of-mass energy of 250 and 350 GeV assuming the CLIC luminosity spectrum for signal HZ events (top), and standard model background (bottom). Figure taken from \cite{Thomson:2015jda}.}
    \label{fig:HZHadronic}
\end{figure}

A concrete example for the exploitation of precise jet energy reconstruction in the context of Higgs physics is the reconstruction of HZ Higgsstrahlung events for hadronic decays of the Z boson. The recoil mass measurements in the HZ process give model-independent access to the total HZ cross section, and with that to the coupling of the Higgs boson to the Z, and thus also allow to constrain invisible Higgs boson decays. In principle, the extension from considering only Z $\rightarrow$ $\mu^+\mu^-,\, e^+e^-$  to the inclusion of hadronic Z boson decays increases the number of HZ signal events by one order of magnitude. The actual improvement achievable by adding these events depends on the level of background in the signal region, which in turn is influenced by the accuracy of the recoil mass reconstruction, which is determined by the jet reconstruction and the beam parameters. Concrete full-simulation studies have been carried out in the context of CLIC and ILC, using the PandoraPFA algorithm \cite{Thomson:2009rp,Marshall:2012ry}. Due to the constrained phase space at 250 GeV close to the threshold of ZH production, there is significant background in the signal region from four-quark final states that end up being reconstructed near the kinematic limit, resulting in only modest improvements in the measurement of the ZH cross section on the order of 20\% compared to leptonic Z decays only, assuming the CLIC luminosity spectrum \cite{Thomson:2015jda}. Nevertheless, \autoref{fig:HZHadronic} (left) demonstrates the capability for a clean reconstruction of the hadronic Z decays and the associated recoil mass. The technique unfolds its full potential at energies of around 350 GeV, where signal and background are well separated, as shown in \autoref{fig:HZHadronic} (right). Here, an improvement of the cross section measurement by a factor of 2.3 is achieved in a CLIC study \cite{Thomson:2015jda,Abramowicz:2016zbo}.

This example demonstrates that the potential for Higgs boson precision measurements profits significantly from increasing performance of the jet energy resolution, resulting in corresponding requirements on the calorimeter and tracker imposed by the PF reconstruction. At the same time, it shows that this performance alone does not guarantee precision, with kinematic boundary conditions and the quality of the association of final state particles to jets also highly relevant. 

In the context of top quark physics, we have already discussed in \autoref{sec:PFA} with the example of \cite{Seidel:2013sqa} that the exploitation of kinematic constraints in the event reconstruction can improve quantities such as the reconstructed invariant mass, and with that can also serve as a means of signal selection. Measurements that use the top quarks as tools to explore physics beyond the standard model impose further requirements on the overall event reconstruction. 
 It is crucial to assess the flavour and the correct association of the jets to the final state partons for measuring asymmetries or searching for CP-violating couplings. Moreover, these will help also increase the sensitivity for the search for FCNC in the top sector, where the current expectation obtained with a traditional analysis approach is comparable to that for the HL-LHC \cite{Abada:2019lih}.

In general, it is important to make use of the fully hadronic six jets final state. A kinematic fitting strategy is necessary to minimise the background contribution, but it is even more crucial to have the correct association of all the decay products to the top and anti-top. Fully exploiting all the event information, such as jet flavour, composition and charge, is necessary to maximise the precision of this measurement. This requires the extension of the PF reconstruction and the flavour tagging strategies outlined in earlier sections with algorithms that provide an accurate determination of the jet charge to enable a unique identification of the top and anti-top quarks. 

With the object reconstruction that is required for precise hadronic energy measurements already well developed, the limitations on global event reconstruction imposed by the uncertainties of the association of individual particles to jets emerge as a key challenge for the full exploitation of the physics potential of the inherently clean $e^+e^-$ collisions. This area has not yet been thoroughly explored, and may hold significant potential for further improvement with more advanced analysis strategies. For Higgs boson physics, strategies for ``colour singlet clustering'' \cite{LC-REP-2013-021,ColorSinglet_TopLC2018,Zhu:2019uve,Ju:2020tbo}, that exploit the physical properties of colour-neutral particles decaying into two hadronic jets, may improve the matching of particles to jets, resulting in a better measurement of the properties of the underlying original partons.  
Such techniques, if proven to be capable of improved precision for the reconstruction of signal final states while also being robust against backgrounds, would improve the measurement of the global event kinematics, and would result in a clearer separation of signal and background. Following this direction may enable further advances, and is expected to lead to new reconstruction paradigms that consider the overall event from the start. Further potential is provided by the increasing power and sophistication of machine learning techniques, which are expected to open new avenues, such as radically different approaches, where the PF candidates are processed directly via ML algorithms to obtain a full event reconstruction \cite{Pata:2021oez,DiBello:2020bas}. Such methods would by-pass the more traditional approaches which rely on the precise reconstruction of the physics objects as an intermediate step. 
This is clearly a very bold approach, yet promising given the clean event topologies at the FCC-ee, as well as the current and future developments in the detector design and the PF algorithm.

\section{Summary}
\label{section:conclusion}

Some of the major priorities of the FCC-ee physics programme, and of the High Energy Physics community in general, is the measurement with unprecedented precision of the Higgs boson and top quark properties. Breakthroughs in multiple areas from detector design, physics object reconstruction and identification, and analysis techniques, are necessary to achieve this goal. It is observed that the detector concepts currently being discussed for future $e^+e^-$ colliders, with high resolution, low mass trackers and highly granular calorimetry, enable object reconstruction and flavour tagging with unprecedented precision. At the same time, this increased precision puts new limitations in the focus, which currently drive the uncertainties in high-multiplicity final states as present in Higgs boson and top quark physics. Novel, more radical approaches to global event reconstruction combining physics-driven algorithms and machine learning techniques may hold significant additional potential for the full exploitation of the excellent experimental conditions provided by FCC-ee.

\bibliographystyle{jhep}
\bibliography{references}

\providecommand{\href}[2]{#2}\begingroup\raggedright\begin{thebibliography}{10}

\bibitem{Abada:2019lih}
{\scshape FCC} collaboration, A.~Abada et~al., \emph{{FCC Physics
  Opportunities}: {Future Circular Collider Conceptual Design Report Volume
  1}}, \href{http://dx.doi.org/10.1140/epjc/s10052-019-6904-3}{\emph{Eur. Phys.
  J. C} {\bfseries 79} (2019) 474}.

\bibitem{Buskulic:1994wz}
{\scshape ALEPH} collaboration, D.~Buskulic et~al., \emph{{Performance of the
  ALEPH detector at LEP}},
  \href{http://dx.doi.org/10.1016/0168-9002(95)00138-7}{\emph{Nucl. Instrum.
  Meth. A} {\bfseries 360} (1995) 481--506}.

\bibitem{Sirunyan:2017ulk}
{\scshape CMS} collaboration, A.~Sirunyan et~al., \emph{{Particle-flow
  reconstruction and global event description with the CMS detector}},
  \href{http://dx.doi.org/10.1088/1748-0221/12/10/P10003}{\emph{JINST}
  {\bfseries 12} (2017) P10003},
  [\href{https://arxiv.org/abs/1706.04965}{{\ttfamily 1706.04965}}].

\bibitem{Thomson:2009rp}
M.~Thomson, \emph{{Particle Flow Calorimetry and the PandoraPFA Algorithm}},
  \href{http://dx.doi.org/10.1016/j.nima.2009.09.009}{\emph{Nucl. Instrum.
  Meth. A} {\bfseries 611} (2009) 25--40},
  [\href{https://arxiv.org/abs/0907.3577}{{\ttfamily 0907.3577}}].

\bibitem{Marshall:2012ry}
J.~Marshall, A.~M\"unnich and M.~Thomson, \emph{{Performance of Particle Flow
  Calorimetry at CLIC}},
  \href{http://dx.doi.org/10.1016/j.nima.2012.10.038}{\emph{Nucl. Instrum.
  Meth. A} {\bfseries 700} (2013) 153--162},
  [\href{https://arxiv.org/abs/1209.4039}{{\ttfamily 1209.4039}}].

\bibitem{ILD:2020qve}
{\scshape ILD Concept Group} collaboration, H.~Abramowicz et~al.,
  \emph{{International Large Detector: Interim Design Report}},
  \href{https://arxiv.org/abs/2003.01116}{{\ttfamily 2003.01116}}.

\bibitem{Sefkow:2015hna}
F.~Sefkow, A.~White, K.~Kawagoe, R.~P\"oschl and J.~Repond, \emph{{Experimental
  Tests of Particle Flow Calorimetry}},
  \href{http://dx.doi.org/10.1103/RevModPhys.88.015003}{\emph{Rev. Mod. Phys.}
  {\bfseries 88} (2016) 015003},
  [\href{https://arxiv.org/abs/1507.05893}{{\ttfamily 1507.05893}}].

\bibitem{Adloff:2012gv}
{\scshape CALICE} collaboration, C.~Adloff et~al., \emph{{Hadronic energy
  resolution of a highly granular scintillator-steel hadron calorimeter using
  software compensation techniques}},
  \href{http://dx.doi.org/10.1088/1748-0221/7/09/P09017}{\emph{JINST}
  {\bfseries 7} (2012) P09017},
  [\href{https://arxiv.org/abs/1207.4210}{{\ttfamily 1207.4210}}].

\bibitem{Tran:2017tgr}
H.~L. Tran, K.~Kr\"uger, F.~Sefkow, S.~Green, J.~Marshall, M.~Thomson et~al.,
  \emph{{Software compensation in Particle Flow reconstruction}},
  \href{http://dx.doi.org/10.1140/epjc/s10052-017-5298-3}{\emph{Eur. Phys. J.
  C} {\bfseries 77} (2017) 698},
  [\href{https://arxiv.org/abs/1705.10363}{{\ttfamily 1705.10363}}].

\bibitem{Bacchetta:2019fmz}
N.~Bacchetta et~al., \emph{{CLD -- A Detector Concept for the FCC-ee}},
  \href{https://arxiv.org/abs/1911.12230}{{\ttfamily 1911.12230}}.

\bibitem{Abramowicz:2016zbo}
H.~Abramowicz et~al., \emph{{Higgs physics at the CLIC
  electron\textendash{}positron linear collider}},
  \href{http://dx.doi.org/10.1140/epjc/s10052-017-4968-5}{\emph{Eur. Phys. J.
  C} {\bfseries 77} (2017) 475},
  [\href{https://arxiv.org/abs/1608.07538}{{\ttfamily 1608.07538}}].

\bibitem{Abramowicz:2018rjq}
{\scshape CLICdp} collaboration, H.~Abramowicz et~al., \emph{{Top-Quark Physics
  at the CLIC Electron-Positron Linear Collider}},
  \href{http://dx.doi.org/10.1007/JHEP11(2019)003}{\emph{JHEP} {\bfseries 11}
  (2019) 003}, [\href{https://arxiv.org/abs/1807.02441}{{\ttfamily
  1807.02441}}].

\bibitem{Seidel:2013sqa}
K.~Seidel, F.~Simon, M.~Tesar and S.~Poss, \emph{{Top quark mass measurements
  at and above threshold at CLIC}},
  \href{http://dx.doi.org/10.1140/epjc/s10052-013-2530-7}{\emph{Eur. Phys. J.
  C} {\bfseries 73} (2013) 2530},
  [\href{https://arxiv.org/abs/1303.3758}{{\ttfamily 1303.3758}}].

\bibitem{Abdallah:2002xm}
{\scshape DELPHI} collaboration, J.~Abdallah et~al., \emph{{b tagging in DELPHI
  at LEP}}, \href{http://dx.doi.org/10.1140/epjc/s2003-01441-8}{\emph{Eur.
  Phys. J. C} {\bfseries 32} (2004) 185--208},
  [\href{https://arxiv.org/abs/hep-ex/0311003}{{\ttfamily hep-ex/0311003}}].

\bibitem{Proriol:1950599}
J.~Proriol, A.~Falvard, P.~Henrard, J.~Jousset and B.~B., \emph{{Tagging B
  quark events in ALEPH with neural networks: comparison of different
  methods}}, {\emph{Int. J. Neural Syst.} {\bfseries 3 Supp.} (Aug, 1991)
  267--270. 26 p}.

\bibitem{CMS-PAS-BTV-11-004}
{\scshape CMS} collaboration, \emph{{b-Jet Identification in the CMS
  Experiment}},  Tech. Rep. CMS-PAS-BTV-11-004, CERN, Geneva, 2012.

\bibitem{ATLAS-CONF-2010-041}
{\scshape ATLAS} collaboration, \emph{{Impact parameter-based b-tagging
  algorithms in the 7 TeV collision data with the ATLAS detector: the
  TrackCounting and JetProb algorithms}},  Tech. Rep. ATLAS-CONF-2010-041,
  CERN, Geneva, Jul, 2010.

\bibitem{ATLAS-CONF-2010-042}
{\scshape ATLAS} collaboration, \emph{{Performance of the ATLAS Secondary
  Vertex b-tagging Algorithm in 7 TeV Collision Data}},  Tech. Rep.
  ATLAS-CONF-2010-042, CERN, Geneva, Jul, 2010.

\bibitem{ATLAS-CONF-2010-070}
{\scshape ATLAS} collaboration, \emph{{Tracking Studies for $b$-tagging with 7
  TeV Collision Data with the ATLAS Detector}},  Tech. Rep.
  ATLAS-CONF-2010-070, CERN, Geneva, Jul, 2010.

\bibitem{ATLAS-CONF-2010-091}
{\scshape ATLAS} collaboration, \emph{{Performance of Impact Parameter-Based
  b-tagging Algorithms with the ATLAS Detector using Proton-Proton Collisions
  at $\sqrt{s}$ = 7 TeV}},  Tech. Rep. ATLAS-CONF-2010-091, CERN, Geneva, Oct,
  2010.

\bibitem{CMS-DP-2017-027}
{\scshape CMS} collaboration, \emph{{New Developments for Jet Substructure
  Reconstruction in CMS}},  Tech. Rep. CERN-CMS-DP-2017-027, Jul, 2017.

\bibitem{CMS:2020poo}
{\scshape CMS} collaboration, A.~M. Sirunyan et~al., \emph{{Identification of
  heavy, energetic, hadronically decaying particles using machine-learning
  techniques}},
  \href{http://dx.doi.org/10.1088/1748-0221/15/06/P06005}{\emph{JINST}
  {\bfseries 15} (2020) P06005},
  [\href{https://arxiv.org/abs/2004.08262}{{\ttfamily 2004.08262}}].

\bibitem{fftforEE}
F.~Bedeschi, L.~Gouskos and M.~Selvaggi, \emph{{private communication, to be
  submitted}},  2021.

\bibitem{Qu:2019gqs}
H.~Qu and L.~Gouskos, \emph{{ParticleNet: Jet Tagging via Particle Clouds}},
  \href{http://dx.doi.org/10.1103/PhysRevD.101.056019}{\emph{Phys. Rev. D}
  {\bfseries 101} (2020) 056019},
  [\href{https://arxiv.org/abs/1902.08570}{{\ttfamily 1902.08570}}].

\bibitem{Abada:2019zxq}
{\scshape FCC} collaboration, A.~Abada et~al., \emph{{FCC-ee: The Lepton
  Collider}: {Future Circular Collider Conceptual Design Report Volume 2}},
  \href{http://dx.doi.org/10.1140/epjst/e2019-900045-4}{\emph{Eur. Phys. J. ST}
  {\bfseries 228} (2019) 261--623}.

\bibitem{deFavereau:2013fsa}
{\scshape DELPHES 3} collaboration, J.~de~Favereau, C.~Delaere, P.~Demin,
  A.~Giammanco, V.~Lema\^\i{}tre, A.~Mertens et~al., \emph{{DELPHES 3, A
  modular framework for fast simulation of a generic collider experiment}},
  \href{http://dx.doi.org/10.1007/JHEP02(2014)057}{\emph{JHEP} {\bfseries 02}
  (2014) 057}, [\href{https://arxiv.org/abs/1307.6346}{{\ttfamily 1307.6346}}].

\bibitem{Hines:2011uf}
{\scshape ATLAS} collaboration, E.~Hines, \emph{{Performance of Particle
  Identification with the ATLAS Transition Radiation Tracker}},  in
  \emph{{Meeting of the APS Division of Particles and Fields}}, 9, 2011.
\newblock \href{https://arxiv.org/abs/1109.5925}{{\ttfamily 1109.5925}}.

\bibitem{Powell:1322666}
A.~Powell, \emph{{Particle Identification at LHCb. Particle ID in LHCb}},
  Tech. Rep. CERN-LHCb-PROC-2011-008, Jan, 2011.

\bibitem{Giammanco:1095044}
A.~Giammanco, \emph{{Particle Identification with Energy Loss in the CMS
  Silicon Strip Tracker}},  Tech. Rep. CMS-NOTE-2008-005, CERN, Geneva, Jun,
  2007.

\bibitem{Thomson:2015jda}
M.~Thomson, \emph{{Model-independent measurement of the e$^{{+}}$ e$^{-}$
  $\rightarrow $ HZ cross section at a future e$^{{+}}$ e$^{-}$ linear collider
  using hadronic Z decays}},
  \href{http://dx.doi.org/10.1140/epjc/s10052-016-3911-5}{\emph{Eur. Phys. J.
  C} {\bfseries 76} (2016) 72},
  [\href{https://arxiv.org/abs/1509.02853}{{\ttfamily 1509.02853}}].

\bibitem{LC-REP-2013-021}
J.~Tian,  and K.~Fujii, \emph{{Summary of Higgs coupling measurements with
  staged running of ILC at 250\,GeV, 500\,GeV and 1\,TeV}},  Tech. Rep.
  LC-REP-2013-021, DESY, Jul, 2013.

\bibitem{ColorSinglet_TopLC2018}
J.~Tian, \emph{{Jet clustering at ILC}}, {\emph{{talk presented at the Workshop
  on Top Physics at the LC 2018, June 4-6, 2018, Sendai, Japan}} }.

\bibitem{Zhu:2019uve}
Y.~Zhu and M.~Ruan, \emph{{Performance study of the full hadronic WW and ZZ
  events\textquoteright{} separation at the CEPC}},
  \href{http://dx.doi.org/10.1140/epjc/s10052-019-6719-2}{\emph{Eur. Phys. J.
  C} {\bfseries 79} (2019) 274}.

\bibitem{Ju:2020tbo}
X.~Ju and B.~Nachman, \emph{{Supervised Jet Clustering with Graph Neural
  Networks for Lorentz Boosted Bosons}},
  \href{http://dx.doi.org/10.1103/PhysRevD.102.075014}{\emph{Phys. Rev. D}
  {\bfseries 102} (2020) 075014},
  [\href{https://arxiv.org/abs/2008.06064}{{\ttfamily 2008.06064}}].

\bibitem{Pata:2021oez}
J.~Pata, J.~Duarte, J.-R. Vlimant, M.~Pierini and M.~Spiropulu, \emph{{MLPF:
  Efficient machine-learned particle-flow reconstruction using graph neural
  networks}},
  \href{http://dx.doi.org/10.1140/epjc/s10052-021-09158-w}{\emph{Eur. Phys. J.
  C} {\bfseries 81} (2021) 381},
  [\href{https://arxiv.org/abs/2101.08578}{{\ttfamily 2101.08578}}].

\bibitem{DiBello:2020bas}
F.~A. Di~Bello, S.~Ganguly, E.~Gross, M.~Kado, M.~Pitt, L.~Santi et~al.,
  \emph{{Towards a Computer Vision Particle Flow}},
  \href{http://dx.doi.org/10.1140/epjc/s10052-021-08897-0}{\emph{Eur. Phys. J.
  C} {\bfseries 81} (2021) 107},
  [\href{https://arxiv.org/abs/2003.08863}{{\ttfamily 2003.08863}}].

\end{thebibliography}\endgroup

\section*{\small Data availability}
{\small \it Raw data were generated at CERN. Derived data supporting the findings of this study are available from the  authors upon request.}

\end{document}